\documentclass{svmult}

\usepackage{makeidx}         
\usepackage{graphicx}        
\usepackage{multicol}        
\usepackage[bottom]{footmisc}

\usepackage{amsmath}         
\usepackage[lft]{floatflt}

\def\LA{\left\langle}
\def\RA{\right\rangle}
\def\LP{\left(}
\def\RP{\right)}
\def\LS{\left[}
\def\RS{\right]}

\makeindex             

\graphicspath{{./fig/}}  

\begin{document}
\title*{Financial time-series analysis: A brief overview}
\author{A. Chakraborti\inst{1}\and
M. Patriarca\inst{2} \and M.S. Santhanam\inst{3}}
\institute{Department of Physics, Banaras Hindu University, Varanasi-221 005, India
\texttt{achakraborti@yahoo.com}
\and Institute of Theoretical Physics, Tartu University, T\"ahe 4, 51010 Tartu, Estonia
\texttt{marco.patriarca@mac.com}
\and Physical Research Laboratory, Navrangpura, Ahmedabad-380 009, 
India
\texttt{santh@prl.ernet.in}
 }
%
%

\maketitle






%
%
%
%
%
%
%
%
%
%

\section{Introduction}

Prices of commodities or assets produce what is called time-series. 
Different kinds of financial time-series have been recorded and studied for 
decades. 
Nowadays, all transactions on a financial market are
recorded, leading to a huge amount of data available, either for free
in the Internet or commercially.
Financial time-series analysis is of great interest to practitioners 
as well as to theoreticians,  for making inferences and predictions.
Furthermore, the stochastic uncertainties inherent in financial time-series 
and the theory needed to deal with them make the subject especially interesting 
not only to economists, but also to statisticians and physicists~\cite{tsay}. 
While it would be a formidable task to make an exhaustive review on the topic,
with this review we try to give a flavor of some of its aspects.

\section{Stochastic methods in time-series analysis} 

The birth of physics as a science is usually associated with the study of mechanical objects moving with negligible fluctuations, such as the motion of planets.
However, this type of systems is not unique, especially at smaller scales where the interaction with the environment and its influence in the form of random fluctuations has to be taken into account.
The main theoretical tool to describe the evolution of such systems is the theory of stochastic processes, which can be formulated in various ways: in terms of a Master equation, Fokker-Planck type equation, random walk model, Langevin equation, or through path integrals. 
Some systems can present unpredictable chaotic behavior due to dynamically generated internal noise.
Either truly stochastic or chaotic in nature, noisy processes represent the rule rather than an exception, not only in condensed matter physics but in many fields such as cosmology, geology, meteorology, ecology, genetics,  sociology, and economics. In fact the first formulation of the random walk model and a stochastic process was given in the framework of an economic study~\cite{LB,Bouchaud2005a}.
In the following we propose and discuss some questions which we consider as possible land-marks in the field of time series analysis.

\subsection{Time-series versus random walk}

What if the time-series were similar to a random walk?
The answer is: It would not be possible to predict future price 
movements using the past price movements or trends. 
Louis Bachelier, who was the first one to investigate this issue
in 1900 \cite{LB}, reached the conclusion 
that ``The mathematical expectation of the speculator is zero'' and
described this condition as a ``fair game.''

\begin{figure}
\begin{center}\resizebox{0.66\textwidth}{!}{\includegraphics{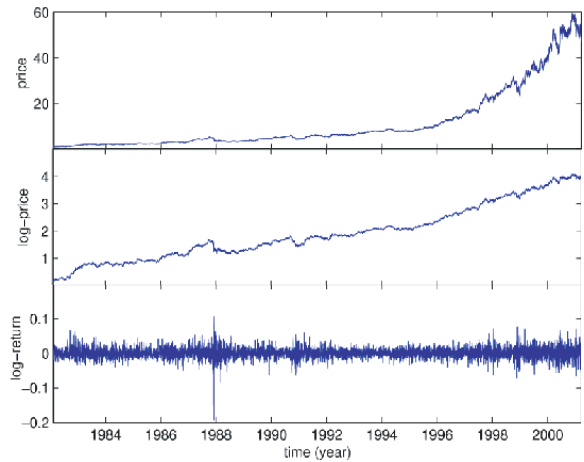} }
\end{center}
\caption{Price in USD (above), log-price (center) and log-return (below) 
plotted versus time for the General
Electric during the period 1982-2000.}
\label{fig:logreturn}
\end{figure}

\begin{figure}[]
\centering
\includegraphics[width=2.5in,angle=0]{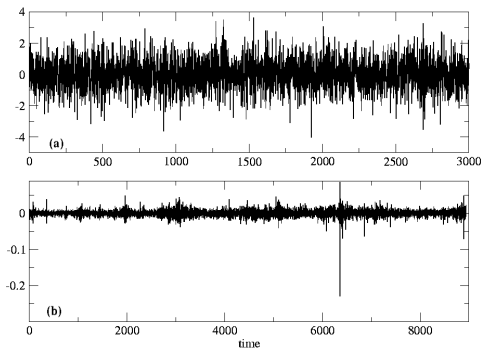}
\caption{
Random time-series, 3000 time steps (above) and Return time-series 
of the S\&P500 stock index, 8938 time steps (below).
}
\label{fig:random}
\end{figure}

In economics, if $P(t)$ is the price of a stock or commodity at time
$t$, then the {}``log-return'' is defined as
$ r_{\tau }(t)=\ln P(t+\tau )-\ln P(t)$,
where $\tau $ is the interval of time. Some statistical features of daily log-return
are illustrated in Fig. \ref{fig:logreturn}, using the price time-series for the General Electric.
The real empirical returns are compared in Fig. \ref{fig:random} 
with a random time-series we generated
using random numbers extracted from a Normal distribution with zero mean 
and unit standard deviation.
If we divide the time-interval $\tau $ into $N$ sub-intervals (of width $\Delta
t$), the total log-return $r_{\tau }(t)$ is by definition the sum
of the log-returns in each sub-interval. 
If the price changes in each sub-interval are independent 
(Fig. \ref{fig:random} above) and identically distributed with a
finite variance, according to the central limit theorem the
cumulative distribution function $F(r_{\tau })$ converges to
a Gaussian (Normal) distribution for large $\tau $.
The Gaussian (Normal) distribution has the following properties: 
(a) the average and most probable change is zero; 
(b) the probability of large fluctuations is very low;
(c) it is a \emph{stable} distribution.
The distribution of returns was first modeled for ``bonds'' 
\cite{LB} as a Normal distribution, 
\[
P(r) = \big[ \sqrt{2\pi }\sigma \big]^{-1} \, \exp (-r^{2}/2\sigma ^{2}) \, ,
\]
where $\sigma ^{2}$ is the variance of the distribution. 

In the classical financial theories Normality had always been assumed, until 
Mandelbrot \cite{Man2} and Fama \cite{key-250} pointed out that the empirical 
return distributions are fundamentally different. 
Namely, they are ``fat-tailed'' and more peaked compared to
the Normal distribution. 
Based on daily prices in
different markets, Mandelbrot and Fama found that $F(r_{\tau })$ was
a stable Levy distribution whose tail decays with an exponent $\alpha \simeq
1.7$.
This result suggested that short-term price changes were
not well-behaved since most statistical properties are not defined
when the variance does not exist. 
Later, using more extensive data,
the decay of the distribution was shown to be fast enough to
provide finite second moment. With time, several other interesting features
of the financial data were unearthed. 

The motive of physicists in analyzing 
financial data has been to find 
common or universal regularities in the complex time-series
(a different approach from those of the economists 
doing traditional statistical analysis of financial data).
The results of their empirical studies on asset price series
show that the apparently random variations of asset prices share some statistical
properties which are interesting, non-trivial, and common for various
assets, markets, and time periods. These are called  ``stylized empirical
facts''. 

\subsection{``Stylized'' facts}

Stylized facts are usually formulated using general
\emph{qualitative} properties of asset returns. Hence,
distinctive characteristics of the individual assets are
not taken into account. 
Below we consider a few ones from Ref.~\cite{Cont1}.


\begin{enumerate}
\item \noindent \textbf{Fat tails}: Large returns asymptotically
follow a power law $F(r_{\tau })\sim \left|r\right|^{-\alpha }$,
with $\alpha >2$. The values $\alpha =3.01\pm 0.03$ and 
$\alpha =2.84\pm 0.12$ are found for the positive
and negative tail respectively~\cite{key-43}. 
An $\alpha > 2$ ensures a well-defined second moment
and excludes stable laws with infinite variance. 
There have been various suggestions for
the form of the distribution: Student's-t (Fig. \ref{fig:sp500daily}),
hyperbolic, normal inverse Gaussian, exponentially truncated
stable, etc. but there no general consensus has been reached yet
%
\begin{figure}
\begin{center}
\resizebox{0.66\textwidth}{!}
{\includegraphics{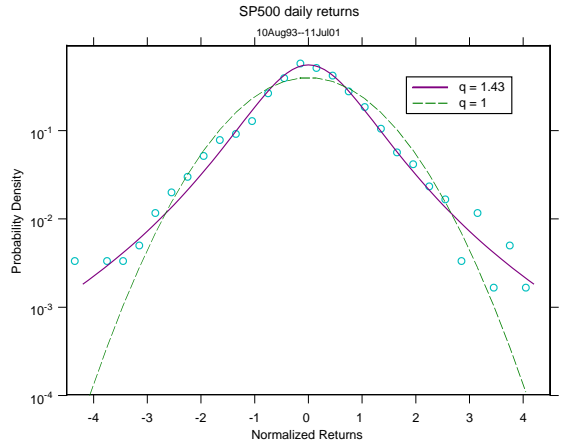} }
\end{center}
\caption{S\&P 500 daily return distribution and normal kernel density estimate.
Distributions of log returns normalized by the sample standard deviation
rising from the demeaned S \& P 500 (circles) and from a Tsallis distribution of index 
$q = 1.43$ (solid line). For comparison, the normal distribution $q = 1$ is shown 
(dashed line). Adapted from Ref.~\cite{cond-mat/0205078}.}
\label{fig:sp500daily}
\end{figure}
\item \noindent \textbf{Aggregational Normality}: As one increases
the time scale over which the returns are calculated, their
distribution 
approaches the Normal form.
The shape is different at
different time scales. The fact that the shape of the distribution
changes with $\tau $ makes it clear that the random process
underlying prices must have non-trivial temporal structure.
\item \noindent \textbf{Absence of linear auto-correlations}: The
auto-correlation of log-returns, $\rho (T)\sim \left\langle
r_{\tau }(t+T)r_{\tau }(t)\right\rangle $, rapidly decays to zero
for $\tau \geq 15$ minutes
\cite{key-44}, which supports the ``efficient market hypothesis'' (EMH),
discussed in Sec.~\ref{emh}. 
When $\tau $ is increased,
weekly and monthly returns exhibit some auto-correlation but the
statistical evidence varies from sample to sample.
\item \noindent \textbf{Volatility clustering}: Price fluctuations
are not identically distributed and the properties of the
distribution, such as the absolute return or variance, change with
time. This is called time-dependent or  ``clustered
volatility''. The volatility measure of absolute returns shows a
positive auto-correlation over a long period of time and decays
roughly as a power-law with an exponent between $0.1$ and $0.3$
\cite{key-44,key-1,key-5}. Therefore high volatility events tend
to cluster in time, large changes tend to be followed by large
changes, and analogously for small changes.
 %
\end{enumerate}

\subsection{The Efficient Market Hypothesis (EMH)}
\label{emh}

A debatable issue in financial econometrics is whether the 
market is ``efficient'' or not.
The ``efficient'' asset market is that in 
which the information 
contained in past prices is instantly, fully and continually reflected in the 
asset's current price. 
The EMH was proposed by Eugene Fama in his 
Ph.D. thesis work in the 1960's, in which he argued that in an 
active market that consists of intelligent and  well-informed investors, 
securities would be fairly priced and reflect all the information available. 
Till date there continues to be disagreement on the degree of market 
efficiency.
The three widely accepted forms of the EMH are:
\begin{itemize}
\item ``Weak'' form: all past market prices and data are fully 
reflected in securities prices and hence technical analysis is of no 
use. 
\item ``Semistrong'' form: all publicly available information is 
fully reflected in securities prices and hence fundamental analysis is 
of no use. 
\item ``Strong'' form: all information is fully reflected in 
securities prices and hence even insider information is of no use.
\end{itemize}

The EMH has provided the basis for much of 
the financial market research. 
In the early 1970's, evidence seemed to be available, supporting the
the EMH: the prices followed a random walk and the 
predictable variations in 
returns, if any, turned out to be statistically insignificant. While most of 
the studies in the 1970's concentrated mainly on predicting prices from past 
prices, studies in the 
1980's looked at the possibility of forecasting based on variables such 
as dividend yield, too, see e.g. Ref.~\cite{FamaFrench}.
Several later studies also looked at things such as the reaction of the stock 
market to the announcement of various events such as takeovers, stock splits, 
etc. 
In general, results from event studies typically showed 
that prices seemed to adjust to new information within a day of the 
announcement of the particular event, an inference that is consistent with the EMH. 
In the 1990's, some studies started looking at the deficiencies of 
asset pricing models. The accumulating evidences suggested that stock 
prices could be predicted with a fair degree of reliability. 
To understand whether 
predictability of returns represented ``rational'' variations in expected 
returns or simply arose as ``irrational'' speculative deviations from 
theoretical values, further studies have been conducted in the recent years. 
Researchers have now discovered several stock market ``anomalies'' that 
seem to contradict the EMH. 
Once an anomaly is discovered, in principle, investors attempting to profit by 
exploiting such an inefficiency should result in the disappearance of the 
anomaly. In fact, many such anomalies that have been 
discovered via back-testing, have subsequently disappeared or proved to be 
impossible to exploit due to high costs of transactions.

We would like to mention the paradoxical nature of efficient markets: 
if every practitioner truly believed that a market was efficient, then the market would 
not have been efficient since no one would have then analyzed the behavior of
the asset prices. 
In fact, efficient markets depend on market participants 
who believe the market is inefficient and trade assets in order to make the 
most of the market inefficiency.

\subsection{Are there any long-time correlations?}

Two of the most important and simple models of 
probability theory and financial econometrics 
are the random walk and the Martingale theory.
They assume that
the future price changes only depend on the past price changes. 
Their main characteristic 
is that the returns are uncorrelated. 
But are they truly uncorrelated or are there 
long-time correlations in the financial time-series? 
This question has been studied especially since it may lead
to deeper insights about the
underlying processes that generate the time-series \cite{stanley}. 

Next we discuss two measures to quantify the long-time correlations, 
and study the strength of trends: the R/S analysis to calculate the Hurst
exponent and the detrended fluctuation analysis 
\cite{vandewalle}.

\subsubsection{Hurst Exponent from R/S Analysis}

In order to measure the strength of trends or ``persistence'' in different 
processes, the rescaled range (R/S) analysis to calculate the Hurst exponent 
can be used. 
One studies the rate of change of the rescaled range 
with the change of the length of time over which measurements are made. 
We divide the time-series $\xi_{t}$ of 
length $T$ into $N$ periods of length 
$\tau$, such that $N\tau=T$. 
For each period $i=1,2,...,N$, containing   
$\tau$ observations, the cumulative deviation is
\begin{equation}
X(\tau) = \sum_{t=(i-1)\tau+1}^{i\tau}\LP \xi_{t}-\LA \xi \RA_{\tau}\RP,
\label{rs1}
\end{equation}
where $\LA \xi \RA_{\tau}$ is the mean within the time-period and is given by 
\begin{equation}
\LA \xi \RA_{\tau}=\frac{1}{\tau}\sum_{t=(i-1)\tau+1}^{i\tau} \xi_{t}.
\end{equation}
The range in the $i$-th time period is given by 
$R(\tau) = \max X(\tau) - \min X(\tau)$,
and the standard deviation is given by
\begin{equation}
S(\tau) =  \LS \frac{1}{\tau} \sum_{t=(i-1)\tau+1}^{i\tau} \LP \xi_{t}-\LA \xi \RA_{\tau}\RP^{2} \RS^{\frac{1}{2}}.
\label{rs3}
\end{equation}
Then $R(\tau)/S(\tau)$ is asymptotically given by a power-law 
\begin{equation}
R(\tau)/S(\tau) = \kappa \tau^{H},
\label{rs4}
\end{equation}
where $\kappa$ is a constant and $H$ the Hurst exponent. In general, 
``persistent'' behavior with fractal properties is characterized
by a Hurst exponent $0.5<H\leq 1$, random behavior by $H=0.5$ and ``anti-persistent'' behavior by $0\leq H< 0.5$. Usually Eq.~(\ref{rs4}) is rewritten in terms of logarithms, 
$\log (R/S)= H \log (\tau)+\log(\kappa)$,
and the Hurst exponent is determined from the slope.

\subsubsection{Detrended Fluctuation Analysis (DFA)}

In the DFA method the time-series $\xi_{t}$ of length $T$ is first divided  
into $N$ non-overlapping periods of length 
$\tau$, such that $N\tau=T$.
In each period $i=1,2,...,N$ the time-series is first fitted through a linear function 
$z_t=at+b$, called the local trend.
Then it is detrended by subtracting the local trend,
in order to compute the fluctuation function,
\begin{equation}
F(\tau) =  \LS \frac{1}{\tau} \sum_{t=(i-1)\tau +1}^{i\tau} \LP \xi_{t}-z_t\RP^{2} \RS^{\frac{1}{2}}.
\label{dfa1}
\end{equation}
The function $F(\tau)$ is re-computed for different box sizes $\tau$ (different scales) 
to obtain the relationship between $F(\tau)$ and $\tau$. A power-law relation 
between $F(\tau)$ and the box size $\tau$, $F(\tau) \sim \tau^{\alpha}$,
indicates the presence of scaling. 
The scaling or ``correlation exponent'' $\alpha$ 
quantifies the correlation properties of the 
signal: if $\alpha=0.5$
the signal is uncorrelated (white noise); 
if $\alpha>0.5$ the signal is anti-correlated; 
if $\alpha<0.5$, there are positive correlations in the signal.

\subsubsection{Comparison of different time-series}

Besides comparing empirical financial time-series with 
randomly generated time-series, here we make the comparison with multivariate 
spatiotemporal time-series drawn from coupled map lattices 
and the multiplicative stochastic process GARCH(1,1)
used to model financial time-series. 

\paragraph{Multivariate spatiotemporal time-series drawn from coupled map 
lattices}

The concept of coupled map lattices (CML) was introduced as a simple
model capable of displaying complex dynamical behavior generic to many
spatiotemporal systems \cite{kan1,kan2}. 
Coupled map lattices are discrete in time 
and space, but have a continuous state space.
By changing the system parameters, one can tune the dynamics toward 
the desired spatial correlation properties, many of them already
studied and reported \cite{kan2}. We consider the class of diffusively 
coupled map lattices in one-dimension, with sites $i=1,2,\dots,n$, of the form
\begin{equation}
y_{t+1}^i 
= (1-\epsilon) f(y_t^i) 
+ \epsilon [ \; f(y_t^{i+1}) + f(y_t^{i-1}) \; ]/2 \, ,
\label{cml}
\end{equation}
where $f(y) = 1 - a y^2$ is the logistic map whose dynamics is
controlled by the parameter $a$ and the parameter $\epsilon$ measures
the coupling strength between nearest-neighbor sites. 
We generally choose periodic boundary conditions, $x(n+1)=x(1)$.
In the numerical computations reported by Chakraborti and Santhanam 
\cite{chakraborti}, a coupled map lattice with
$n=500$ was iterated 
starting from random initial conditions, for $p=5 \times 10^7$
time steps, after discarding $10^5$ transient iterates.
As the parameters $a$ and $\epsilon$ are varied, the spatiotemporal map
displays various dynamical features like frozen random patterns, 
pattern selection, space-time intermittency, and spatiotemporal chaos 
\cite{kan2}. In order to study the coupled map lattice dynamics found in the
regime of spatiotemporal chaos, where correlations are
known to decay rather quickly as a function of the lattice site, 
the parameters were chosen as $a=1.97$ and $\epsilon=0.4$.

\paragraph{Multiplicative stochastic process GARCH(1,1)}

Considerable interest has been in the application of ARCH/GARCH models
to financial time-series, which exhibit periods of unusually 
large volatility followed by periods of relative tranquility. The assumption of
constant variance or ``homoskedasticity'' is inappropriate in such 
circumstances. A stochastic process with auto-regressional conditional 
``heteroskedasticity'' (ARCH) is actually a stochastic process with 
``non-constant variances conditional on the past but constant unconditional 
variances'' \cite{engle}.
The ARCH($p$) process is defined by the equation
\begin{equation}
\sigma_t^2=\alpha_0+\alpha_1 x_{t-1}^{2}+...+\alpha_p x_{t-p}^{2} \, ,
\label{archp}
\end{equation}
where the $\{\alpha_0,\alpha_1,...\alpha_p\}$ are positive parameters and $x_{t}$ is a 
random variable with zero mean and variance $\sigma_t^2$, characterized by
a conditional probability distribution 
function $f_t(x)$, which may be chosen as Gaussian. The nature of the memory
of the variance $\sigma_t^2$ is determined by the parameter $p$.

The generalized ARCH process GARCH($p,q$) was introduced 
by Bollerslev \cite{bollerslev} and is defined by the equation
\begin{equation}
\sigma_t^2=\alpha_0+\alpha_1 x_{t-1}^{2}+...+\alpha_q x_{t-q}^{2}+\beta_{1}\sigma_{t-1}^{2}+...+\beta_{p}\sigma_{t-p}^{2} \, ,
\label{garchpq}
\end{equation}
where $\{\beta_{1},...,\beta_{p}\}$ are additional control parameters.

The simplest GARCH process is the GARCH(1,1) process, with Gaussian 
conditional probability distribution function 
, 
\begin{equation}
\sigma_t^2=\alpha_0+\alpha_1 x_{t-1}^{2}+\beta_{1}\sigma_{t-1}^{2} \, .
\label{garch11}
\end{equation}
%
%
The random variable $x_{t}$ can be written in term of $\sigma_t$ defining
$x_{t}\equiv\eta_t\sigma_t$,
where $\eta_t$ is a random Gaussian process with zero mean and unit variance.
One can rewrite Eq. \ref{garch11} as a random multiplicative process
\begin{equation}
\sigma_t^2=\alpha_0+(\alpha_1 \eta_{t-1}^{2}+\beta_{1})\sigma_{t-1}^{2} \, .
\label{garch12}
\end{equation}

\subsubsection{DFA analysis of auto-correlation function of absolute returns}

\begin{figure}
\includegraphics[width=2.3in,angle=0]{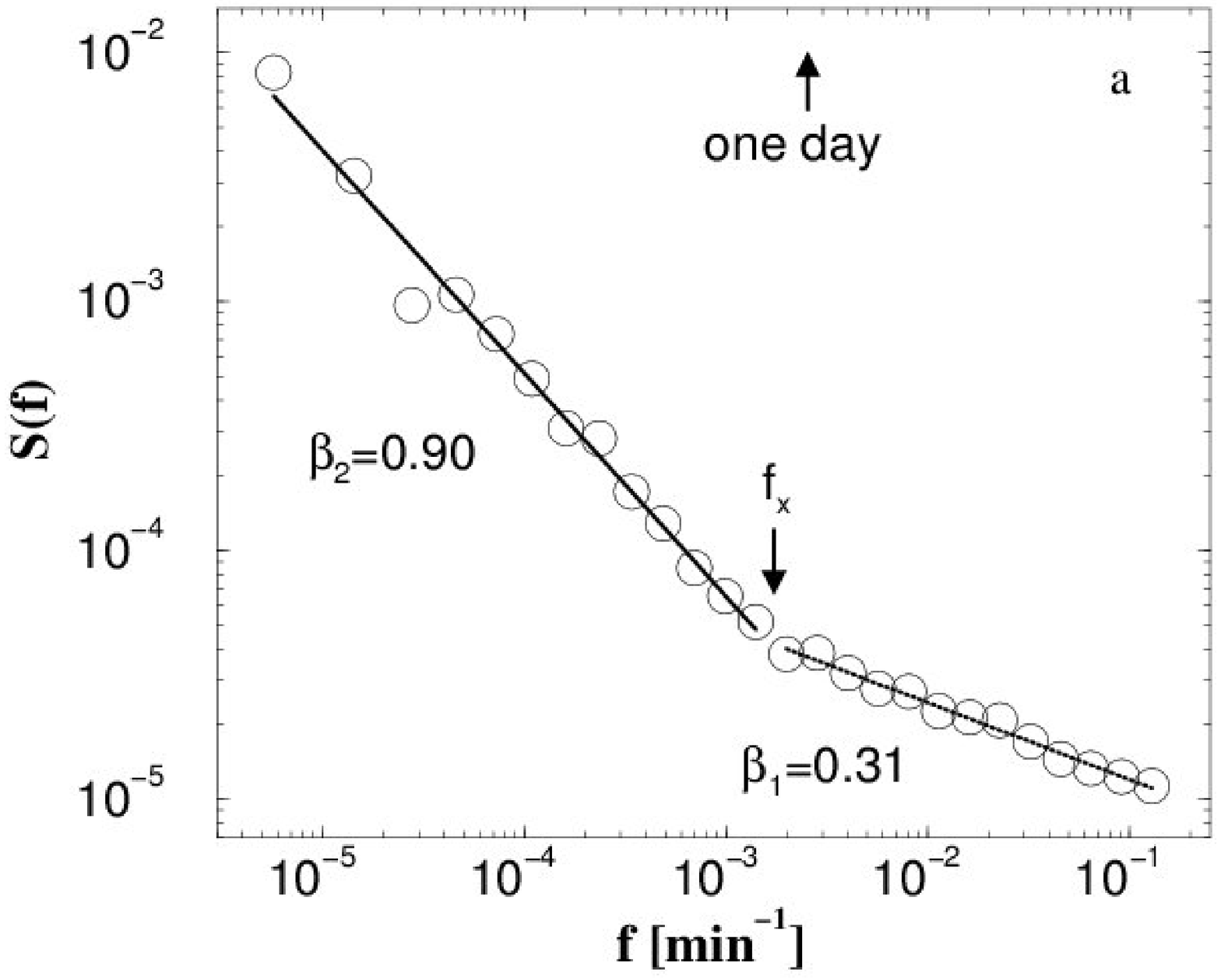}
\includegraphics[width=2.3in,angle=0]{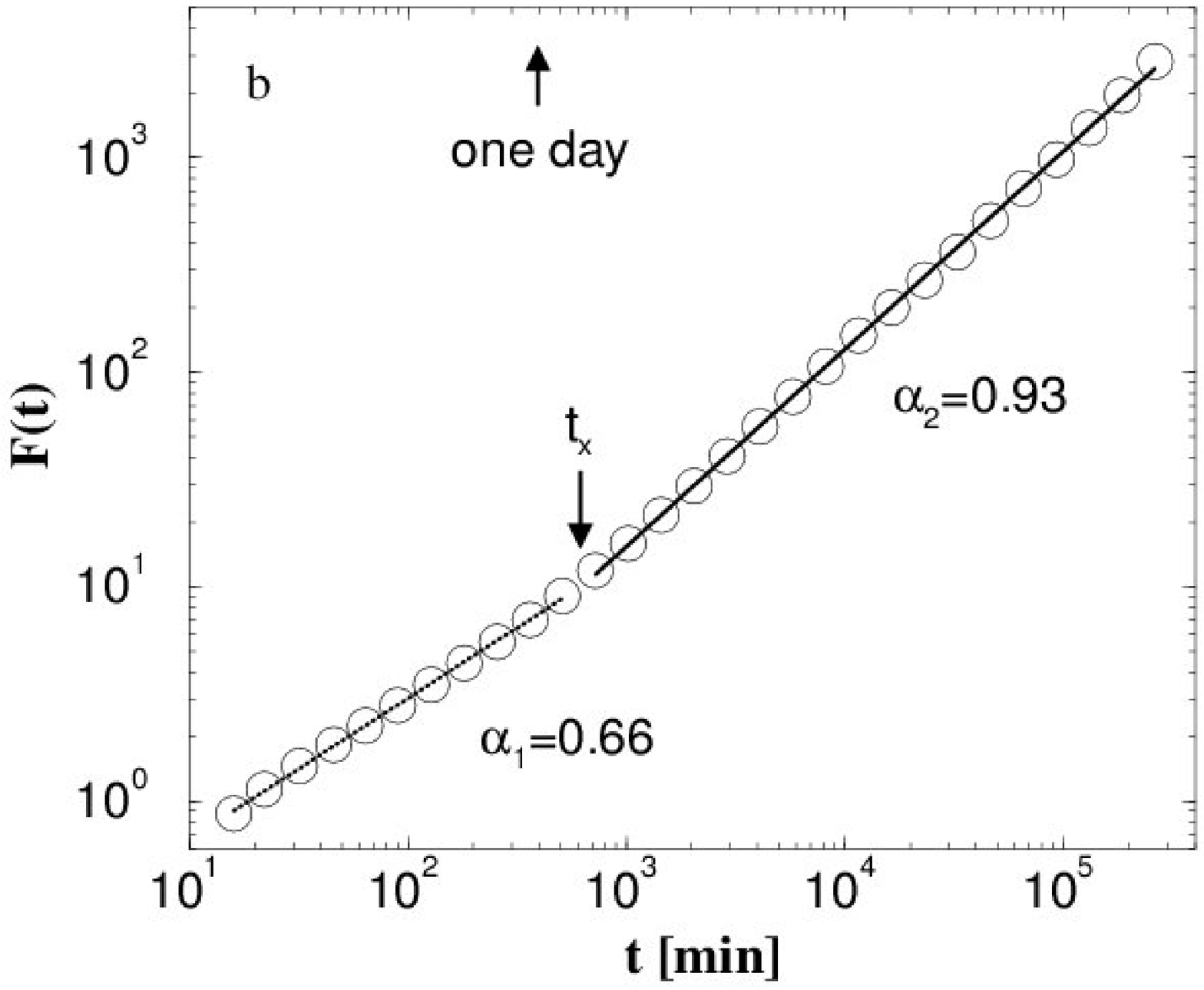}
\caption{Power spectrum analysis (left) and detrended fluctuation analysis (right) of auto-correlation function of absolute returns, from Ref.~\cite{key-1}.}
\label{fig:stanley}
\end{figure}

The analysis of financial correlations was done 
in 1997 by the group of H.E. Stanley \cite{key-1}. The correlation
function of the financial indices of the New York stock exchange and
the S\&P 500  
between January, 1984 and December, 1996
were analyzed at one minute intervals. 
The study confirmed that the 
auto-correlation function of the returns fell off exponentially but the 
absolute value of the returns did not. 
Correlations of the absolute values of the index returns
could be described through two different power laws, with crossover
time $t_{\times }\approx 600$ minutes, corresponding to $1.5$ trading
days. 
Results from power spectrum analysis and DFA analysis were found to be
consistent. 
The power spectrum analysis of Fig. \ref{fig:stanley} yielded
exponents $\beta _{1}=0.31$ and $\beta _{2}=0.90$ for $f>f_{\times }$
and $f<f_{\times }$, respectively. 
This is consistent with the result that 
$\alpha =(1+\beta )/2$ and $t_{\times }\approx 1/f_{\times }$, 
as obtained from detrended fluctuation analysis with exponents $\alpha _{1}=0.66$
and $\alpha _{2}=0.93$ for $t<t_{\times }$ and $t>t_{\times }$,
respectively.
%
%
%
\subsubsection{Numerical Comparison}
\begin{figure}[ht]
\includegraphics[width=1.9in,angle=-90]{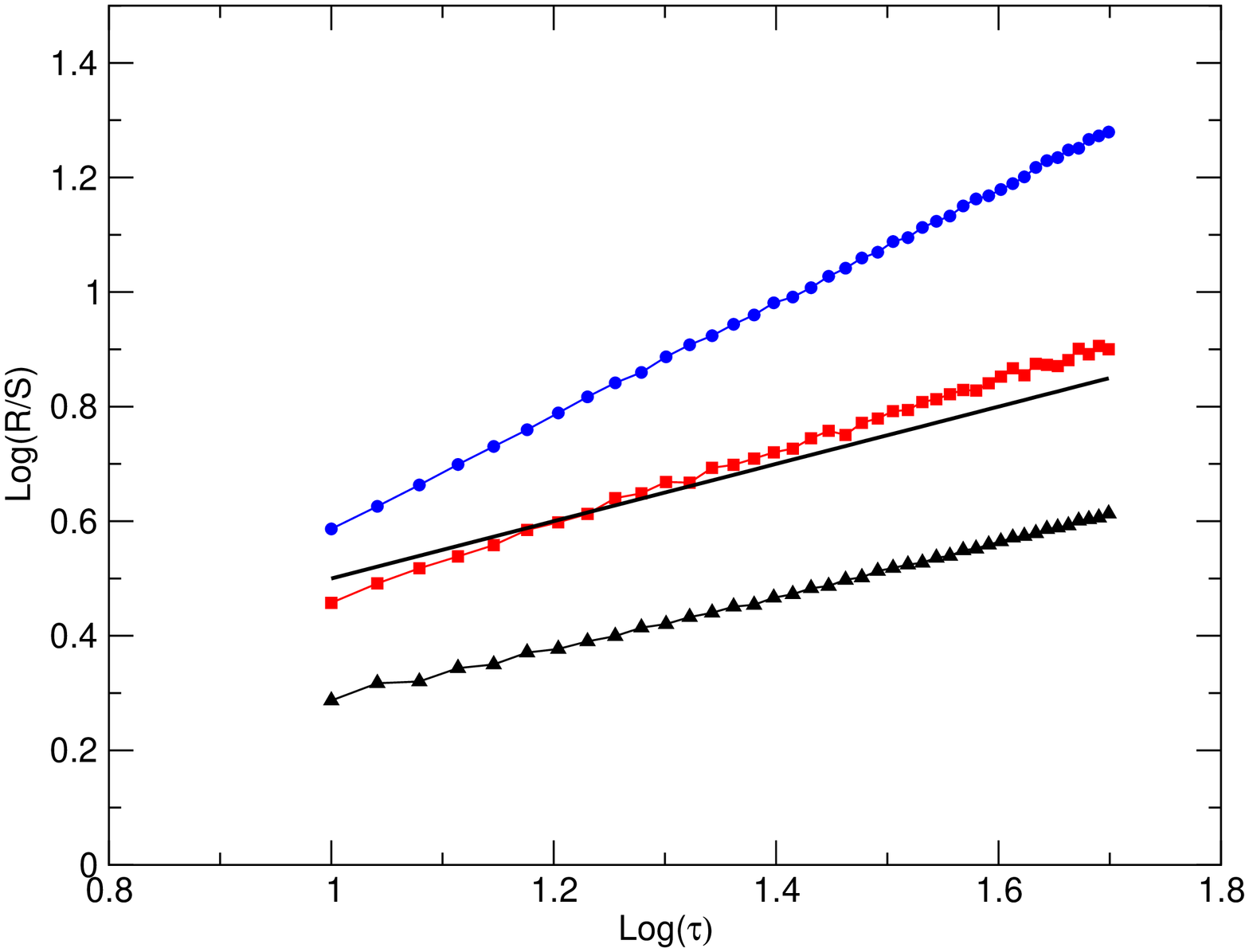}
\includegraphics[width=1.9in,angle=-90]{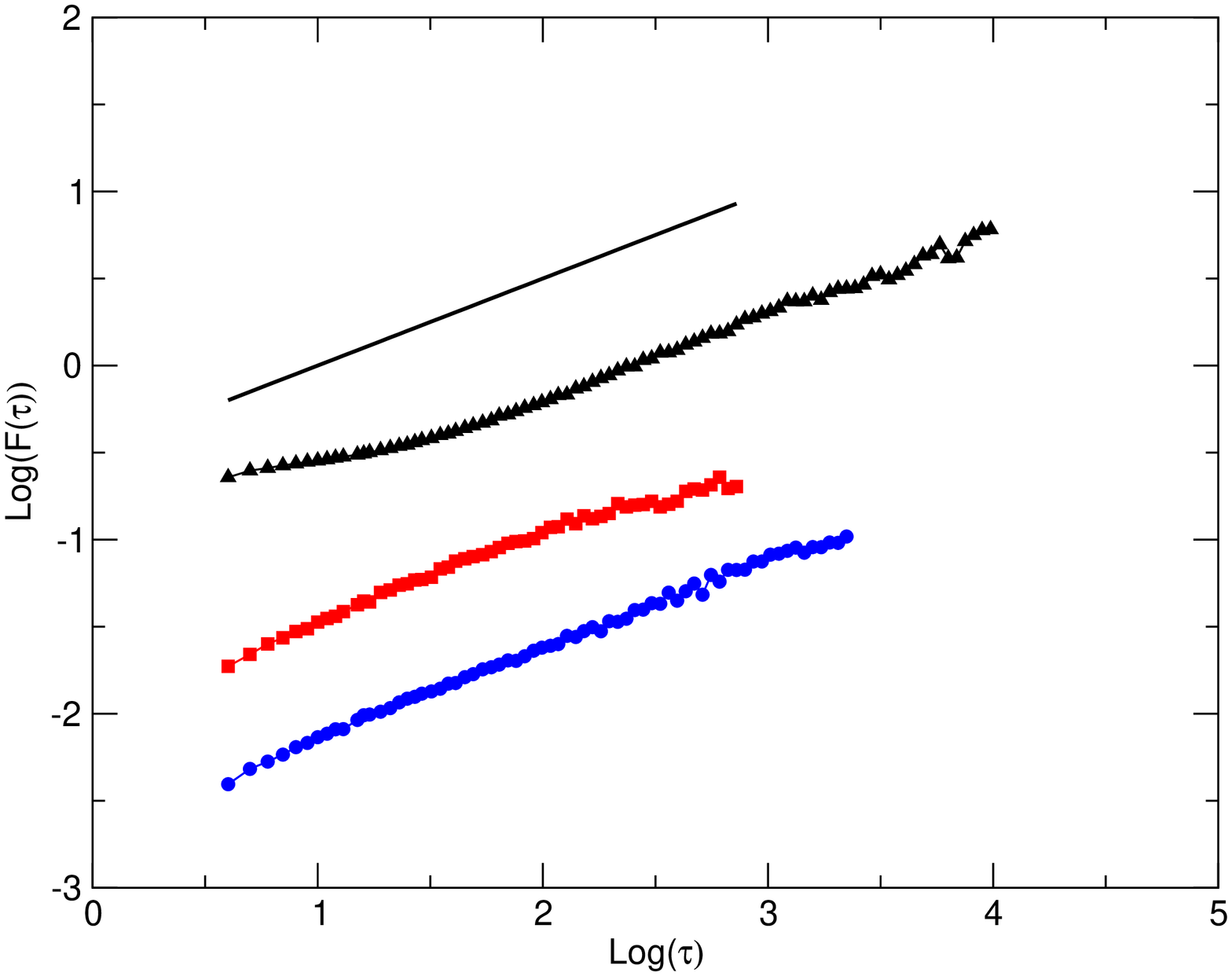}
\caption{
R/S (left) and DFA (right) analyses: Random time-series, 3000 time steps (black solid line); multivariate spatiotemporal time-series drawn from coupled map lattices with parameters 
$a=1.97$ and $\epsilon=0.4$, 3000 time steps (black filled up-triangles); multiplicative stochastic 
process GARCH(1,1) with parameters $\alpha_0 = 0.00023$, $\alpha_1 = 0.09$ and
$\beta_0 = 0.01$, 3000 time steps (red filled squares); Return time-series of the 
S\&P500 stock index, 8938 time steps (blue filled circles).
}
\label{hurstdfa}
\end{figure}

In order to provide an illustrative example,
in Fig. \ref{hurstdfa} a comparison among various analysis techniques and process is presented,
while the values of the exponents of the Hurst and DFA analyzes are listed in Table \ref{exponents}.
For the numerical computations reported by Chakraborti and Santhanam 
\cite{chakraborti}, the parameter values chosen were
$\alpha_0 = 0.00023$, $\alpha_1 = 0.09$ and
$\beta_0 = 0.01$.
%
%
\begin{table}[ht]
\caption{\label{exponents}Hurst and DFA exponents.}
\centering \begin{tabular}{|c|c|c|}
\hline
Process&Hurst exponent&DFA exponent\\
\hline
\hline
Random&0.50&0.50\\
\hline
Chaotic
(CML)&0.46&0.48\\
\hline
GARCH(1,1)&0.63&0.51\\
\hline
Financial Returns&0.99&0.51\\
\hline
\end{tabular}
\end{table}

%
%
%
\section{Random Matrix methods in time-series analysis}

The R/S and the detrended fluctuation analysis considered 
in the previous section are suitable for analyzing univariate data.
Since the stock-market data are essentially \emph{multivariate}
time-series data, it is worth constructing 
a correlation matrix to study its spectra and contrasting it 
with random multivariate data from coupled map lattice. 
Empirical spectra of correlation matrices, drawn from
time-series data, are known to follow  mostly random matrix theory (RMT) \cite{gopi}.

\subsection{Correlation matrix and Eigenvalue density}

\subsubsection{Correlation matrix}
\paragraph{Financial Correlation matrix}

If there are $N$ assets with a price $P_{i}(t)$ for asset $i$ at
time $t$, the logarithmic return of stock $i$ is 
$r_{i}(t)=\ln P_{i}(t)-\ln P_{i}(t-1)$.
A sequence of such values for a give period of time forms 
the return vector $\boldsymbol{r}_{i}$. 
In order to characterize the synchronous
time evolution of stocks, one defines 
the equal time correlation coefficients
between stocks $i$ and $j$,
\begin{equation}
\rho _{ij}
=
\Big[
\langle \boldsymbol r_{i}\boldsymbol r_{j}\rangle -\langle \boldsymbol r_{i}\rangle \langle \boldsymbol r_{j}\rangle 
\Big] \Big/
\sqrt{[\langle \boldsymbol r_{i}^{2}\rangle -\langle \boldsymbol r_{i}\rangle ^{2}][\langle \boldsymbol r_{j}^{2}\rangle -\langle \boldsymbol r_{j}\rangle ^{2}]}
\, ,
\label{corrcoeff}
\end{equation}
where $\left\langle ...\right\rangle $ indicates a time
average over the trading days included in the return vectors. 
The correlation coefficients $\rho _{ij}$ form an $N\times N$ matrix, 
with $-1\leq \rho _{ij}\leq 1$.
If $\rho _{ij}=1$, the stock price changes are completely correlated;
if $\rho _{ij}=0$, the stock price changes are uncorrelated and if
$\rho _{ij}=-1$, then the stock price changes are completely anti-correlated.
%
%
\paragraph{Correlation matrix from spatiotemporal series from coupled map lattices}

Consider a time-series of the form $z'(x,t)$, where $x\!=\!1,2,\dots,n$
and $t\!=\!1,2,\dots,p$ denote the discretized space and time.
In this way, the time-series at every spatial point is treated as a
different variable. 
We define 
\begin{equation}
  z(x,t) = \big[ z'(x,t) - \langle z'(x)\rangle \big] \big/ {\sigma(x)} \, ,
\end{equation}
as the normalized variable, with the brackets $\langle . \rangle$
representing a temporal average and $\sigma(x)$
the standard deviation of $z'$ at position $x$.
Then, the equal-time cross-correlation matrix 
can be written as
\begin{equation}
S_{x,x'} = 
\langle z(x,t) \, z(x',t) \rangle \, ,
~~~~~~ x, x' = 1,2,\dots,n \, .
\end{equation}
This correlation matrix is symmetric by construction. In addition,
a large class of processes is translationally invariant and the
correlation matrix will possess the corresponding symmetry. 
We use this property for our correlation models in the context of coupled map lattices.
In time-series analysis, the averages $\langle . \rangle$ have to be
replaced by estimates obtained from finite samples. 
We use the maximum likelihood estimates, i.e., 
$\langle a(t) \rangle \approx \frac{1}{p}\sum_{t=1}^p a(t)$. 
These estimates contain statistical uncertainties which disappear for $p\to\infty$.
Ideally we require $p \gg n$ to have reasonably correct correlation estimates.

\subsubsection{Eigenvalue Density}

The interpretation of the 
spectra of empirical correlation matrices should be done carefully 
in order to distinguish between system specific signatures and universal
features. 
The former ones express themselves in a smoothed level density, whereas 
the latter ones are usually represented by the fluctuations on top of such a smooth 
curve. 
In time-series analysis, matrix elements are not only prone to
uncertainties such as measurement noise on the time-series data, but also
to the statistical fluctuations due to finite sample effects. 
When characterizing time
series data in terms of RMT, we are not interested in
these 
sources of fluctuations, which are present on every data set,
but we want to identify the significant features which would be
shared, in principle, by an ``infinite'' amount of data without measurement noise.
The eigenfunctions of the correlation matrices
constructed from such empirical time-series carry the information
contained in the original time-series data in a ``graded'' manner and
provide a compact representation for it. 
Thus, by applying an approach based on RMT, 
we try to identify non-random components of the
correlation matrix spectra as deviations from RMT predictions 
\cite{gopi}.

\begin{figure}[ht]
\includegraphics[width=1.8in,angle=-90]{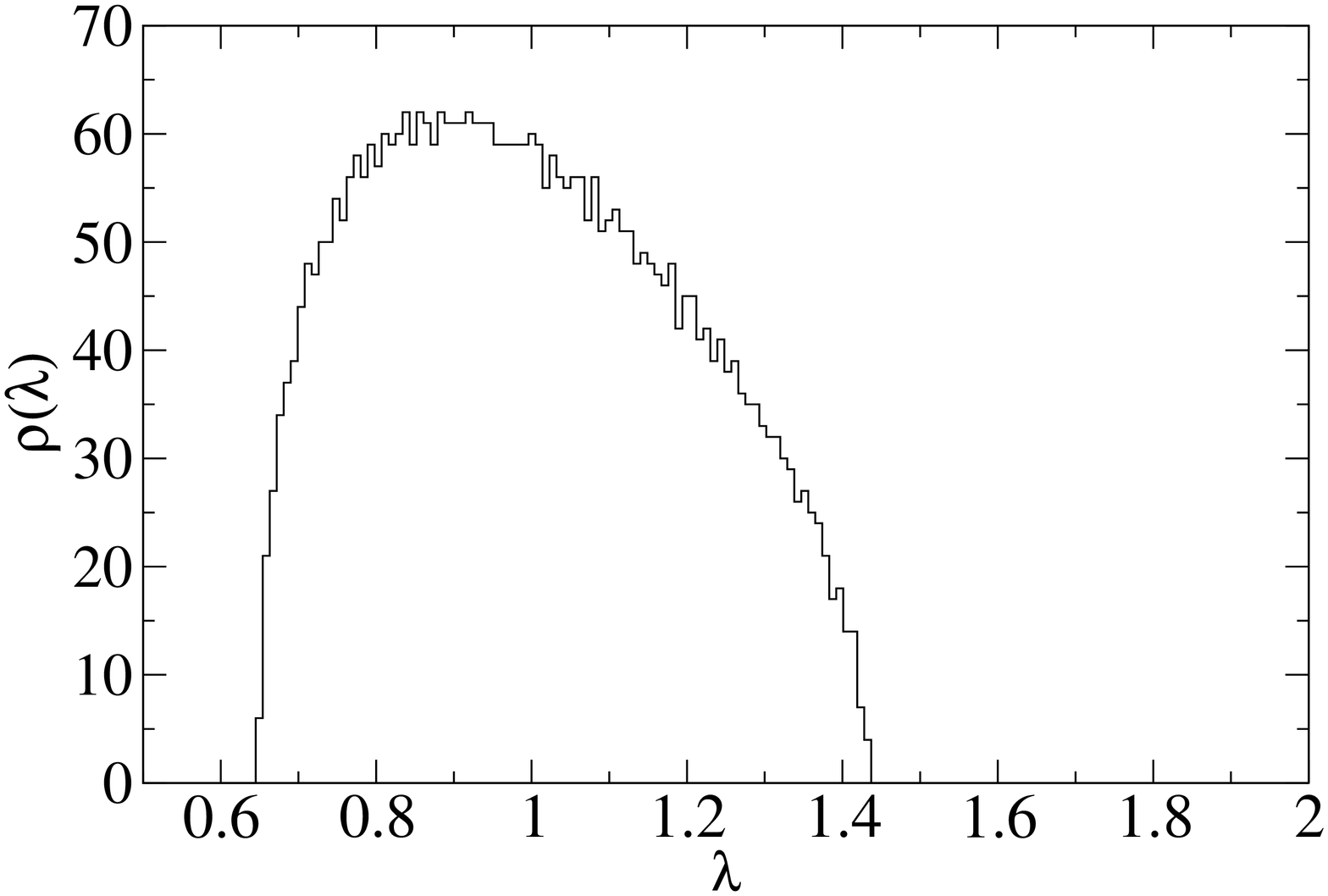}
\includegraphics[width=1.8in,angle=-90]{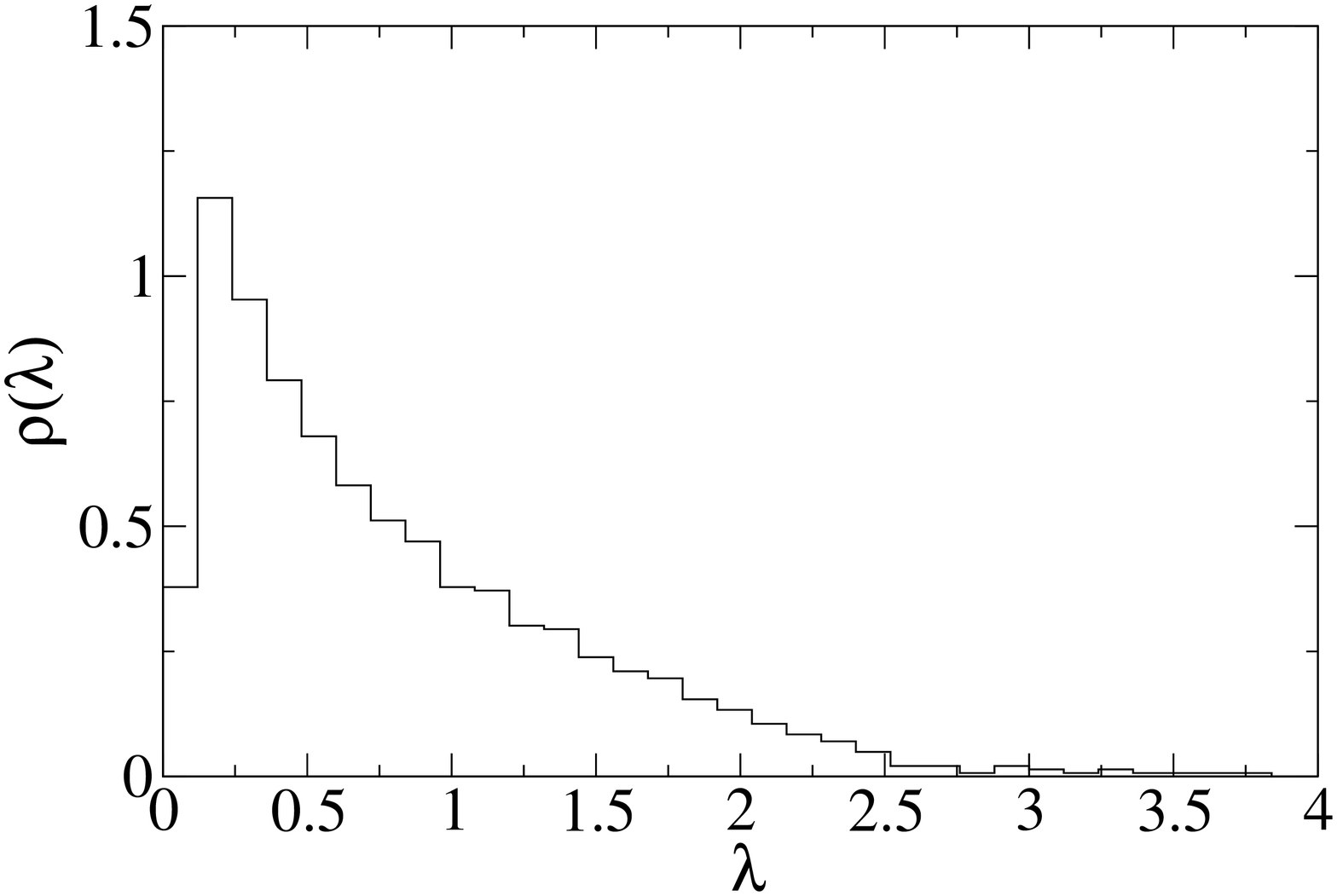}
\caption{ Spectral density for
multivariate spatiotemporal time-series drawn from
coupled map lattices (left) and eigenvalue density for the
return time-series of the S\&P500 stock market data,
8938 time steps (right).
}
\label{fig3}
\end{figure}

We now consider the eigenvalue density, studied in applications
of RMT methods to time-series correlations.
Let ${\mathcal N}(\lambda)$ be the integrated eigenvalue density, giving
the number of eigenvalues smaller than a given $\lambda$.
The eigenvalue or level density, 
$\rho(\lambda) = d\mathcal N(\lambda)/d\lambda$,
can be obtained assuming a random correlation matrix \cite{mitra}.
Results are found to be in good agreement with the empirical time-series data from stock 
market fluctuations \cite{plerou}. 
From RMT considerations, the eigenvalue
density for random correlations is given by
\begin{equation}
\rho_{rmt}(\lambda) 
= 
[Q/(2 \pi \lambda)] \sqrt{{(\lambda_{max}-\lambda})({\lambda-\lambda_{min}})} \, .
\label{rho}
\end{equation}
Here $Q\!=\!N/T$ is the ratio of the number of variables to the length of each 
time-series, while $\lambda_{min} = 1 + 1/Q - 2 \sqrt{1/Q}$ and 
$\lambda_{max} = 1 + 1/Q + 2 \sqrt{1/Q}$ represent the minimum and maximum 
eigenvalues of the random correlation matrix.
The presence of correlations in the empirical correlation matrix produces
a violation of this form of eigenvalue density, for a certain number of dominant
eigenvalues, often corresponding to system specific information in the data.
As examples, Fig. \ref{fig3} shows the eigenvalue densities for S\&P500 data and
for the chaotic data from coupled map lattice are shown: the curves
are qualitatively different from the form of Eq.~(\ref{rho}). 

\subsection{Earlier estimates and studies using Random Matrix Theory (RMT)}

\begin{figure}
\begin{center}\resizebox{0.48\textwidth}{!}
{\includegraphics{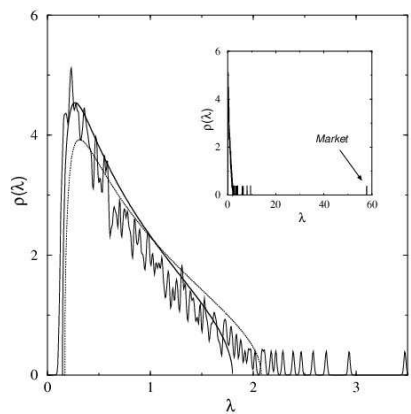} }
\end{center}
\caption{Eigenvalue spectrum of the correlation matrices, adapted from Ref.~\cite{key-4}.}
\label{fig:eigensp}
\end{figure}

Laloux et al. \cite{key-4} showed that results from RMT 
were useful to understand the statistical structure of
the empirical correlation matrices appearing in the study of price
fluctuations. The empirical determination of a correlation matrix
is a difficult task. If one considers $N$ assets, the correlation
matrix contains $N(N-1)/2$ mathematically independent elements, which
must be determined from $N$ time-series of length $T$. If $T$ is
not very large compared to $N$, then generally the determination
of the covariances is noisy, and therefore the empirical correlation
matrix is to a large extent random. The smallest eigenvalues of the
matrix are the most sensitive to this ``noise''. But the eigenvectors
corresponding to these smallest eigenvalues determine the minimum
risk portfolios in Markowitz's theory. It is thus important to distinguish
``signal'' from ``noise'' or, in other words, to extract the eigenvectors
and eigenvalues of the correlation matrix, containing real information
(which is important for risk control), from those which do not contain
any useful information and are unstable in time. It is useful to compare
the properties of an empirical correlation matrix to a  ``null hypothesis''
--- a random matrix which arises for example from a finite time-series
of strictly uncorrelated assets. Deviations from the random matrix
case might then suggest the presence of true information. The main
result of the study was a remarkable agreement between theoretical
predictions, based on the assumption that the correlation matrix is
random, and empirical data concerning the density of eigenvalues. 
This is shown in Fig. \ref{fig:eigensp} for the time-series of the different
stocks of the S\&P 500 (or other stock markets).

Cross-correlations in financial data were also studied by Plerou et
al. \cite{key-2}, who analyzed price fluctuations 
of different stocks through RMT. 
Using two large databases, they calculated cross-correlation matrices of
returns constructed from: (i) 30-min returns of 1000 US stocks for
the 
period 1994--95; (ii) 30-min returns of 881 US stocks for
the 
period 1996--97; (iii) 1-day returns of 422 US stocks
for the 
period 1962--96. They tested the statistics of the 
eigenvalues $\lambda _{i}$ of cross-correlation matrices against a  ``null
hypothesis'' and found that a majority of the eigenvalues of the
cross-correlation matrices were within the RMT bounds 
$(\lambda _{min},\lambda _{max})$ defined above
for random correlation matrices. 
Furthermore, they analyzed
the eigenvalues of the cross-correlation matrices within the RMT bound
for universal properties of random matrices and found good agreement
with the results for the Gaussian orthogonal ensemble (GOE) of random matrices,
implying a large degree of randomness in the measured cross-correlation
coefficients. It was found that: (i) the distribution of eigenvector
components, for the eigenvectors corresponding to the eigenvalues outside
the RMT bound, displayed systematic deviations from the RMT prediction;
(ii) such ``deviating eigenvectors'' were stable in time; 
(iii) the largest eigenvalue corresponded to
an influence common to all stocks;
(iv) the remaining deviating eigenvectors showed distinct groups, 
whose identities corresponded
to conventionally-identified business sectors.

\section{Approximate Entropy method in time-series analysis}

The Approximate Entropy (ApEn) method is an information theory-based estimate 
of the complexity of a time series introduced by S. Pincus~\cite{Pincus1991a}, 
formally based on the evaluation of joint probabilities, in a way similar to 
the entropy of Eckmann and Ruelle.
The original motivation and main feature, however, was not to characterize an 
underlying chaotic dynamics, rather to provide a robust model-independent 
measure of the randomness of a time series of real data, possibly --- as it is 
usually in practical cases --- from a limited data set affected by a superimposed noise.
ApEn has been used by now to analyze data obtained from 
very different sources, such as digits of irrational and transcendental 
numbers, hormone levels, clinical 
cardiovascular time-series, anesthesia 
depth, EEG time-series, and 
respiration in various conditions.

Given a sequence of $N$ numbers $\{u(j)\} = \{ u(1), u(2), \dots, u(N)\}$, with equally spaced times $t_{j+1}-t_j \equiv \Delta t = \mathrm{const}$, one first extracts the sequences with embedding dimension $m$, i.e., $x(i) = \{u(i), u(i+1), \dots, u(i + m - 1)\}$, with $1 \le i \le N - m + 1$.
The ApEn is then computed as
\begin{equation}
  \mathrm{ApEn} = \Phi^m(r) - \Phi^{m+1}(r) \, ,
\end{equation}
where $r$ is a real number representing a threshold distance between series,
 and the quantity $\Phi^m(r)$ is defined as
\begin{equation}
  \Phi^m(r) = \langle \ln[C_i^m(r)] \rangle 
  = \sum_{i = 1}^{N-m+1}\ln[C_i^m(r)] / (N - m + 1) \, .
\end{equation}
Here $C_i^m(r)$ is the probability that the series $x(i)$ is closer to a 
generic series $x(j)$ ($j \le N - m + 1$) than the threshold $r$,
\begin{equation}
  C_i^m(r) = \mathcal{N}[d(i,j) \le r] / (N - m + 1) \, ,
\end{equation}
with $\mathcal{N}[d(i,j) \le r]$ the number of sequences $x(j)$ close to $x(i)$ less than $r$.
As definition of distance between two sequences, the maximum difference (in 
modulus) between the respective elements is used,
\begin{equation}
  d(i,j) = \max_{k = 1, 2, \dots, m}(|u(j + k - 1) - u(i + k - 1)|) \, .
\end{equation}
Quoting Pincus and Kalman~\cite{Pincus2004a}, ``\dots ApEn measures the 
logarithmic frequency that runs of patterns that are close (within $r$) for $m$
 contiguous observations remain close (within the same tolerance width $r$) on 
the next incremental comparison''.
Comparisons are intended to be done at fixed $m$ and $r$, the general 
ApEn($m$,$r$) being in fact a family of parameters. 

In economics, the ApEn method has been shown to be a reliable estimate of the 
efficiency of market~\cite{Pincus1991a,Pincus2004a,Pincus1996a} and has been 
applied to various economically relevant events.
For instance, the ApEn computed for the S\&P 500 index has shown a drastic 
increase in the two-week period preceding the stock market crash of 1987.
Just before the Asian crisis of November 1997, the ApEn computed for the 
Hong Kong's Hang Seng index, from 1992 to 1998, assumes its highest values. 
More recently, a broader investigation carried out for various countries 
through the ApEn by Oh, Kim, and Eom, revealed a systematic difference 
between the efficiencies of the markets between the period before and after 
the the Asian crisis~\cite{Oh2006a}.




\printindex

\begin{thebibliography}{99.}


\bibitem{tsay}
R.S. Tsay, 
Analysis of Financial Time Series,
John Wiley, New York (2002).
%


\bibitem{LB}
L. Bachelier,
Theorie de la speculation. 
Annales Scientifiques de l'Ecole Normale Superieure, Suppl. 3, No. 1017, 21-86 (1900). 
English translation by A. J. Boness in: P. Cootner (Ed.), 
The Random Character of Stock Market Prices, Page 17, MIT, Cambridge, MA, (1967).

\bibitem{Bouchaud2005a}
J.-P. Bouchaud,
{CHAOS} {\bf 15}, 026104 (2005).

\bibitem{Man2}
B.B. Mandelbrot, 
J. Business \textbf{36}, 394 (1963).

\bibitem{key-250}
E. Fama, 
J. Business {\bf 38}, 34 (1965).

\bibitem{Cont1}
R. Cont, 
Quant. Fin. \textbf{1}, 223 (2001).

\bibitem{cond-mat/0205078}
L. Borland, 
{\tt arxiv:cond-mat/0205078}.

\bibitem{key-43}
P. Gopikrishnan, M. Meyer, L.A.N. Amaral, H.E. Stanley, 
Eur. Phys. J. B {\bf 3}, 139 (1998).

\bibitem{key-44}
R. Cont, M. Potters, J.-P. Bouchaud, in: 
B. Dubrulle, F. Graner and D. Sornette (Eds.),
Scale Invariance and Beyond 
(Proc. CNRS Workshop on Scale Invariance, Les Houches, 1997), 
Springer, Berlin (1997).

\bibitem{key-1}
Y. Liu, P. Cizeau, M. Meyer, C.-K. Peng, H.E. Stanley, 
Physica A \textbf{245}, 437 (1997); {\tt arxiv:cond-mat/9706021}.

\bibitem{key-5}
P. Cizeau, Y. Liu, M. Meyer, C.-K. Peng, H.E. Stanley, 
Physica A \textbf{245}, 441 (1997).

\bibitem{FamaFrench}
E.F. Fama, K.R. French, 
J. Fin. Economics {\bf 22}, 3 (1988). 

\bibitem{stanley}
R.N. Mantegna, H.E. Stanley, 
An Introduction to Econophysics,
Cambridge University Press, New York (2000).

\bibitem{vandewalle}
N. Vandewalle, M. Ausloos, 
Physica A {\bf 246}, 454 (1997).
C.-K. Peng, S.V. Buldyrev, S. Havlin, M. Simons, H.E. Stanley, A.L. Goldberger,
Phys. Rev. E {\bf 49}, 1685 (1994). 
Y. Liu, P. Gopikrishnan, P. Cizeau, M. Meyer, C.-K.. Peng, H.E. Stanley,
Phys. Rev. E {\bf 60}, 1390 (1999).
M. Beben, A. Orlowski, 
Eur. Phys. J. B {\bf 20}, 527 (2001).
A. Sarkar, P. Barat, 
Physica A {\bf 364}, 362 (2006);
{\tt arxiv:physics/0504038}.
P. Norouzzadeh, B. Rahmani,
Physica A {\bf 367}, 328 (2006);
D. Wilcox, T. Gebbie, 
{\tt arxiv:cond-mat/0404416}. 





\bibitem{kan1} 
See K. Kaneko (Ed.), Theory and Applications of Coupled Map Lattices, Wiley, New York (1993), and in particular the contribution of R. Kapral.

\bibitem{kan2} 
K. Kaneko, 
Physica D {\bf 34}, 1 (1989).

\bibitem{chakraborti}
A. Chakraborti, M.S. Santhanam, 
Int. J. Mod. Phys. C {\bf 16}, 1733 (2005).

\bibitem{engle}
R.F. Engle, 
Econometrica {\bf 50}, 987 (1982).

\bibitem{bollerslev}
T. Bollerslev, 
J. Econometrics {\bf 31}, 307 (1986).




\bibitem{gopi} 
P. Gopikrishnan, B. Rosenow, V. Plerou, H.E. Stanley,
Phys. Rev. E {\bf 64}, 035106 (2001).






\bibitem{mitra} 
A.M. Sengupta, P.P. Mitra, 
Phys. Rev. E {\bf 60}, 3389 (1999).

\bibitem{plerou} 
V. Plerou, P. Gopikrishnan, B. Rosenow, L.A.N. Amaral, T. Guhr, H.E. Stanley, 
Phys. Rev. E {\bf 65}, 066126 (2002).

\bibitem{key-4}
L. Laloux, P. Cizeau, J.-P. Bouchaud, M. Potters,
Phys. Rev. Lett. {\bf 83}, 1467 (1999);
{\tt arxiv:cond-mat/9810255}.


\bibitem{key-2}
V. Plerou, P. Gopikrishnan, B. Rosenow, L.A.N. Amaral, H.E. Stanley, 
Phys. Rev. Lett. {\bf 83}, 1471 (1999);
{\tt arxiv:cond-mat/9902283}.
V. Plerou, P. Gopikrishnan, B. Rosenow, L.A.N. Amaral, T. Guhr, H.E. Stanley, 
Phys. Rev. E {\bf 65}, 066126 (2002); {\tt arxiv:cond-mat/0108023}.




\bibitem{Pincus1991a}
S.~M. Pincus,
Proc. Nati. Acad. Sci. USA {\bf 88}, 2297 (1991).

\bibitem{Pincus2004a}
S.~Pincus, R.E. Kalman,
Proc. Nati. Acad. Sci. USA {\bf 101}, 13709 (2004).

\bibitem{Pincus1996a}
S.~Pincus, B.H. Singer,
Proc. Nati. Acad. Sci. USA {\bf 93}, 2083 (1996).





\bibitem{Oh2006a}
G.~Oh, S.~Kim, C.~Eom,
Market efficiency in foreign exchange markets,
Physica A, in press.

\end{thebibliography}
\end{document}